\documentclass[onecolumn]{aa}  

\usepackage{graphicx}
\usepackage{txfonts}
\begin{document}

   \title{Naked emergence of an anti-Hale active region }

   \subtitle{I. Overall evolution and magnetic properties}
   \author{Jincheng Wang\inst{1,2,3} 
   \and Xiaoli Yan\inst{1,3} 
   \and Defang Kong\inst{1,3} 
   \and Zhike Xue\inst{1,3}
   \and Liheng Yang\inst{1,3} 
   \and Qiaoling Li\inst{1,4}
   \and Yan Zhang\inst{1,4}
   \and Hao Li\inst{5,6}
    }
   \institute{Yunnan Observatories, Chinese Academy of Sciences, Kunming Yunnan 650216, PR China\\
         \email{wangjincheng@ynao.ac.cn}
         \and
             CAS Key Laboratory of Solar Activity, National Astronomical Observatories, Beijing 100012, PR China
            \and
            Center for Astronomical Mega-Science, Chinese Academy of Sciences, 20A Datun Road, Chaoyang District, Beijing, 100012, PR China
            \and
            University of Chinese Academy of Sciences, Yuquan Road, Shijingshan Block Beijing 100049, PR China
            \and
            Instituto de Astrofisica de Canarias, E-38205 La Laguna, Tenerife, Spain
            \and
            Universidad de La Laguna, Dpto. Astrofisica, E-38206 La Laguna, Tenerife, Spain
             }

   \date{Received March 1, 2021; accepted **** **, ****}
   
   \abstract
   { }
   {In order to understand the emergence of the active region, we investigate the emerging process and magnetic properties of a naked anti-Hale active region during the period between August 24 to 25, 2018.}
   {Using the data from Helioseismic and Magnetic Imager on board the Soar Dynamic Observatory and the New Vacuum Solar Telescope, we calculated different evolving parameters (such as pole separation, tilt angle) and magnetic parameters (such as vertical electric current, force-free parameter, relative magnetic helicity) during the emergence of the active region. With these calculated parameters and some reasonable assumptions, we use two different methods to estimate the twist of the active region.}
   {The magnetic flux and pole separation continue increasing while the tilt angle exhibits a decreasing pattern during the emergence of the active region. The increase of the pole separation is mainly contributed as a result of the enhancement in the longitude direction. A power-law relationship between pole separation and total flux is found during the emergence of the active region. On the other hand, it is found that both the positive and negative electric currents increased equivalently and the average flux-weighted force-free parameter $\tilde \alpha$ remains  almost consistently positive, on the order of $\sim$ 10$^{-8}$ m$^{-1}$. The relative magnetic helicity is mainly contributed by the shear term, while the relative magnetic helicity injection flux of the shear term changes its sign at the latter stage of the emergence. The twist number of the whole active region remains on the order of 10$^{-1}$ turns during the emergence of the active region.}
    {We find that the magnetic flux tube with low twist also could emerge into the solar atmosphere.}
    
 \maketitle
 \section{Introduction}
 It is widely known that the solar atmosphere is dominated by magnetic fields and almost all the activities taking place in the atmosphere are associated with the evolution of magnetic fields (e.g.,\citealp{tho11,wie14,van15,cai19}). Therefore, the question of the origin of magnetic fields is the most significant aspect in understanding  the nature of solar activities. The observational sunspots and active regions (ARs) are thought to be born of the emerging magnetic flux, which is the concentration of the magnetic fields \citep{van15}. On the other hand, many solar activities are triggered by newly emerging magnetic flux, such as coronal mass ejections (CMEs), flares, formations or eruptions of the filaments, jets, and small-scale bursts (e.g., \citealp{fey95,che00,iso17,yan17,che18,wan18,wan19,you18,tia18}). Such an emerging magnetic field not only plays a vital role in understanding active regions and sunspots, but it also has a great influence on many of the activities in taking place in the solar atmosphere.
 
Flux-emerging regions (FERs), or general ARs, are thought to be formed by the rise of flux tubes from the convection zone to the solar surface \citep{van15}. \cite{fu16} found that about half of the investigated emerging active regions show a two-step emergence pattern that indicates that the flux emergence rate was relatively low at the early phase, undergoing a significant enhancement in the next phase. According to the numerical simulations, a twisted flux tube can emerge from the convection zone into the corona, passing through the photosphere and chromosphere as a result of the magnetic buoyancy or Parker instability, which leads to the formation of an $\Omega$-shaped flux tube \citep{lek96,fan01,mag03}. However, some studies have suggested that not all the twisted flux tubes embedded in the subsurface could emerge into the upper atmosphere. \cite{mur06} simulated different emerging flux tubes with different twist values. They found that the highly twisted flux tube could  easily emerge into the upper atmosphere, while the weakly twisted flux tube fails to do so. \cite{tor11} also obtained a similar result that only the flux tube with high twist (q$H_0$ $>$ 0.1, $q=B_\theta/rB_l$, $H_0$=170 km) can be launched into the corona and the low twisted flux tube could hardly emerge into the photosphere due to the influence of the horizontal expansion. However, it is rare to observe a high twisted flux tube emerging from the subsurface. Several researchers have provided some observational evidence to ascertain the emergence of a small twisted flux rope \citep{oka09,yan17}. \cite{poi15} analyzed the ``magnetic tongue '' of 41 emerging bipolar active regions with a new systematic and user-independent method, deducing that flux ropes embedded in the subsurface before the emergence have a low amount of twist and that highly twisted flux tubes are relatively rare. The discrepancy between the numerical simulations and the observations raises a significant question regarding how much twist the flux tube needs for its emergence. 

Magnetic fields emerging from the Sun's interior would carry intrinsic information about the magnetic field in the subsurface to the upper atmosphere. Emerging magnetic fields play a key role in the transportation of the magnetic energy and helicity. When a flux tube emerges into the solar atmosphere from the convective zone (CZ), it not only brings magnetic flux that contributes to the formations of active regions or sunspots but that also injects magnetic energy and magnetic helicity into the solar atmosphere \citep{dem09}. Using a three-dimensional MHD numerical simulation, \cite{mag03} investigated the injection process of magnetic energy and relative magnetic helicity into the upper atmosphere. They found that both the magnetic energy and relative magnetic helicity are contributed by the emergence term during the early phase and the shear term during the latter phase. \cite{liu12} studied the injections of magnetic energy and helicity in two emerging active regions, and they found that rates of energy injection from both the emergence term and the shear term have a consistent evolution in phase during the entire process of flux emergence. They also derived that the injection of magnetic helicity and magnetic energy in the corona are mainly contributed by the shear term and the emergence term, respectively. A similar result was also obtained by \cite{vem15}. However, with the analysis of 28 emerging active regions, \cite{liu14} certified again that the shear term is dominant in the helicity injection, however, both terms, namely, the shear and emergence terms contribute approximately equivalent  energy to the solar atmosphere. On the other hand, \cite{vem17} even found that opposite magnetic helicity was injected into the corona during the emergence of the active region and suggested that it may be caused by the emergence of a flux rope with different sign of twist between its apex and its legs. Investigations of the magnetic helicity or energy injection during the emergence of magnetic flux tube provide a good window for us to improve the understanding of the non-potential origination of emerging magnetic flux and a potential way to explore the nature of the magnetic flux embedded in the subsurface.

In general, the leading sunspots of a bipolar AR in the northern hemisphere have the opposite polarity with regard to leading spots in the southern hemisphere and the leading magnetic polarities alternate in successive sunspot cycle, known as Hale's law \citep{hal25}. This feature can be interpreted well in the models of ``dynamo theory'' (e.g., \citealp{par55,bab61,lei64,wan91,cho95}). Shearing motion by differential rotation, called the $\Omega$-effect, plays a key role in these models \citep{kra80,dik01}. According to the Hale's law, the leading spots of bipolar ARs should be negative  and the following spots should be positive in the northern hemisphere in the solar cycle 24. The ARs with the reverse polarities in leading spots and following spots are considered as anti-Hale ARs. People found that a few ARs could violate Hale's law and the percentage of anti-Hale ranges from a few to 10\%  \citep{wan89,li12,mcc14,li18,zhu20}. \cite{li18} found that  the distribution of tilt angles and magnetic fluxes between Hale and anti-Hale is different while the pole separation of anti-Hale ARs is smaller than Hale ones. The anti-Hale and Hale ARs exhibit similar latitudinal distributions \citep{mcc14,li18}. On the other hand, anti-Hale ARs may hide the imprint of fundamental processes in solar interior. With the clue of the increase of the percentage of anti-Hale regions during the solar minimum, \cite{sok15} concluded that small-scale dynamo is active in the solar interior.  Based on a surface flux transport simulation and observational data, \cite{jia15} proposed that bigger anti-Hale ARs emerging around low latitudes are responsible for the weak polar fields that affect the activity of the subsequent cycle. Thus, studies of anti-Hale ARs may shed light on our understanding of solar magnetic field evolution and supply a clue to probe the performing mechanism of magnetic field. 

In this paper, we focus on the evolution and the magnetic characteristics in a strictly naked emerging anti-Hale active region NOAA 12720. The evolution of the active region and variations of different related-magnetic parameters during its emergence are studied in detail. The sections of this paper are organized as follows. The observations and methods are described in Section \ref{sec:obser methods}, the results are given in Section \ref{sec:results}, and the summary and discussions are presented in Section 4.
\section{Observations and methods} \label{sec:obser methods}
\subsection{Observations}
In this study,  data from Helioseismic and Magnetic Imager (HMI: \citealp{sch12}) on board the Solar Dynamic Observatory\footnote{\url{https://sdo.gsfc.nasa.gov}} (SDO: \citealp{pes12}) are used to analyze the magnetic properties of an emerging Active Region NOAA 12720 during the period from 16:12 UT on August 24 to 12:00 UT on August 25, 2018. The SDO/HMI provides continuous full-disk measurement of the Stokes vector of the photospheric Fe $\rm{I}$ 6173 $\rm{\AA}$ line about every 135 s. The HMI science team had pipelined the process of retrieving vector field information from filtergrams and derived the available data (hmi.sharp.cea.720s) for solar researchers \citep{hoe14}. The pipeline procedure involves inversion of stokes vectors using the Very Fast Inversion of the Stokes Vector algorithm (VFSIV: \citealp{bor11,cen14}) based on the Milne-Eddington atmospheric model and removing the 180$\degr$ azimuthal ambiguity using the minimum energy method \citep{met94,lek09}. Space Weather HMI Active Region Patch (SHARP) products provide the data of AR's line-of-sight and vector magnetic field, continuum intensity, Doppler velocity, with a pixel scale of about 0.5 $\arcsec$ and a cadence of 12 minutes \citep{hoe14}. These data are remapped using Lambert (cylindrical equal area method \citep{cal02}) projection centered on the midpoint of the active region, which is tracked at the Carrington rotation rate. The continuum intensity images from SHARP permit us to investigate the evolution of the active region, whereas the vector magnetograms (VMs) from SHARP allow us to analyze the magnetic properties of the active region. Moreover, the high-resolution TiO and H$\alpha$ images from the New Vacuum Solar Telescope\footnote{\url{http://fso.ynao.ac.cn}} (NVST:\citealp{liu14b,yan20}) are also utilized to exhibit the photospheric {and chromospheric} details. The TiO images have a pixel size of 0.\arcsec052 and a cadence of 30 s. The field of view (FOV) of TiO images is 120\arcsec $\times$ 100\arcsec. The field of view (FOV) of H$\alpha$ images is 150\arcsec $\times$ 150\arcsec, with a 45s cadence and a spatial resolution of 0\arcsec165 per pixel. These data are calibrated from Level 0 to Level 1 with dark current subtracted and flat field corrected, and then speckle masking method was used to reconstruct the calibrated images from Level 1 to Level 1+  \citep{xia16}.
\subsection{Methods}\label{methods}
\subsubsection{Pole separation and tilt angle}
In order to determine the pole separation of the active region, we first need to determine the locations of each polarity. For each moment, the centroids of positive and negative polarities is determined by using the flux-weighted method as follows:
\begin{equation}
(\tilde{x}_{\pm},\tilde{y}_{\pm})=(\frac{\Sigma xB_{\pm}}{\Sigma B_{\pm}}, \frac{\Sigma yB_{\pm}}{\Sigma B_{\pm}}),\label{eq1}
\end{equation}
in which $B_{\pm}$ is the magnetic field strength in each pixel and ($x$, $y$) is the position coordinate of each pixel. The subscript signs  ``+'' and ``-'' denote positive and negative polarities, respectively. As the centroids $(\tilde{x}_{\pm},\tilde{y}_{\pm})$ of the positive and negative polarities have been derived, the pole separation is calculated by using the following equation:
\begin{equation}
 d=\sqrt{(\tilde{x}_+-\tilde{x}_-)^2+(\tilde{y}_+-\tilde{y}_-)^2}.\label{eq2}
 \end{equation}
On the other hand, we define the tilt angle as the angle between the pole separation vector and the solar east. Thus, the tilt angle can be calculated by using the following equation:
\begin{equation}
\vartheta = -\arctan (\frac{\tilde{y}_+-\tilde{y}_-}{\tilde{x}_+-\tilde{x}_-}).\label{eq3}
\end{equation}
\subsubsection{Magnetic flux, vertical electric current, and force-free parameter $\alpha$}
Magnetic fluxes are integrated the positive and negative polarity magnetic fields by their area, respectively. Positive and negative magnetic fluxes are calculated by using the following equations:
\begin{eqnarray}
\varphi_{zp} = \int_{S_{ph}} B_{z}(x,y)_+dS, \nonumber\\
\varphi_{zn} = \int_{S_{ph}} |B_{z}(x,y)_-|dS, \label{eq4}
\end{eqnarray}
where $B_{z}(x,y)_+$/$B_{z}(x,y)_-$ is the vertical positive/negative magnetic field and the S denotes the integrated area with positive or negative magnetic field.

According to Ampere's law, the electric current density in the corona can be derived by the following equation:
\begin{equation}
\textbf{\emph J}=\frac{c}{4\pi}(\triangledown \times \textbf{\emph B}),\label{eq5}
\end{equation}
in which $c$ is the speed of the light and $\emph{\textbf{B}}$ denotes the magnetic field vector in the corona. According to Eq. \ref{eq5}, by neglecting the effect of the electric displacement current, the electric current density component perpendicular to the solar surface ($J_z$) can be calculated by the following equation:
\begin{equation}
J_z(x,y)=\frac{c}{4\pi}(\triangledown \times \emph{\textbf{B}})_z=\frac{c}{4\pi}(\frac{\partial B_y(x,y)}{\partial x}-\frac{\partial B_x(x,y)}{\partial y}),\label{eq6}
\end{equation}
where $B_x$ and $B_y$ are the two perpendicular components of the transverse magnetic fields. With the vector magnetic fields from SDO/HMI and Eq. \ref{eq6}, the distribution of vertical electric current density in the photosphere can be approximated by using a five-point stencil method. Therefore, the vertical current can be integrated by the vertical current density in the active region. The positive and negative currents calculated by using the following equations:
\begin{eqnarray}
  I_{zp}&=&\int _{S_{ph}} J_{z}(x,y)_+ dS, \nonumber \\
  I_{zn}&=&\int_{S_{ph}} |J_{z}(x,y)_-|dS,\label{eq7}
\end{eqnarray}
where $J_{z}(x,y)_+$/$J_{z}(x,y)_-$ is the vertical positive or negative current density and S denotes the integrated area. On the other hand, we also calculate the current neutralization ratio ($\Re$)\citep{tor14,liu17}:
\begin{equation}
 \Re = \frac{I_{ret}}{I_{dir}}, \label{eq8}
\end{equation}
where $I_{ret}$ and $I_{dir}$ are the return and direct currents, respectively. They are express as 
\begin{eqnarray}
 I_{ret}=\int _{S_{ph}} \kappa_{ret}(x,y) J_{z}(x,y)dS, \nonumber\\
 I_{dir}=\int_{S_{ph}}\kappa_{dir}(x,y)J_{z}(x,y)dS,
\end{eqnarray}
with
\begin{eqnarray}
\kappa_{ret}=\left\{
            \begin{array}{cl}
            0& $   $\text{if}$ $J_z(x,y)B_z(x,y)<0 \\
            1& $   $\text{otherwise,}
            \end{array}
            \right.\nonumber
\end{eqnarray}
\begin{eqnarray}
\kappa_{dir}=\left\{
            \begin{array}{cl}
            0& $   $\text {if}$ $J_z(x,y)B_z(x,y)>0 \\
            1& $   $\text{otherwise.}
            \end{array}
            \right.\nonumber
\end{eqnarray}
The current neutralization ratio ($\Re$) is an important parameter in evaluating the equilibrium of the AR system, which is related to flares and CMEs \citep{mel91,liu17,ava20}.

Under the assumption of force-free field, electric currents are parallel or anti-parallel to the magnetic field lines. The electric current could be expressed as:
\begin{equation}
J=\alpha B,\label{10}
\end{equation}
in which $\alpha$ is called force-free parameter, and it is constant along each field line. The parameter $\alpha$ could usually be used to represent a measure of magnetic twist in an AR \citep{see90,pev95,hag04}. A calculable quantity ($\alpha_z=\frac{J_z(x,y)}{B_z(x,y)}$) is a commonly used proxy of the twist at the photospheric level \citep{pev94}. Here, we calculate an averaged flux-weighted force-free parameter $\tilde{\alpha}$ over the whole AR in the photosphere, based on the following equation \citep{hag04,kut19}:
\begin{equation}
  \tilde{\alpha}=\frac{\int j_z(x,y)B_z(x,y)dS}{\int B_z(x,y)^2dS},\label{eq11}
\end{equation}
in which $B_z$ is the vertical magnetic field. S denotes the area of the whole active region.
\subsubsection{Magnetic helicity injection flux}
Relative magnetic helicity (for simplicity, we use magnetic helicity to refer to relative magnetic helicity in this paper) can be transported from the solar interior to the corona by the new emerging fluxes and the various motions of magnetic flux in the photosphere. The magnetic helicity injection flux across a surface S is expressed as \citep{ber84,dem03}:
\begin{equation}
\dot H = 2\int_S(\textbf{\emph A}_p\cdot \textbf{\emph B}_t)V_{\bot n}dS - 2\int_S(\textbf{\emph A}_p\cdot \textbf{\emph V}_{\bot t})B_n dS,\label{eq12}
\end{equation}
where $\textbf{\emph A}_p$ is the vector potential of the potential field $\text{\emph B}_p$, $\textbf{\emph B}_t$, and $B_n$ are the tangential and normal components of the magnetic field, respectively. $\textbf{\emph V}_{\bot t}$ and $\textbf{\emph V}_{\bot n}$ are the tangential and normal components of the velocity perpendicular to the magnetic field lines, respectively. The velocity ($\textbf{\emph V}_\bot$) is further corrected by removing the irrelevant field aligned plasma flow as $\textbf{\emph V}_\bot=\textbf{\emph V}-\frac{\textbf{\emph V}\cdot \textbf{\emph B}}{\textbf{\emph B}^2}\textbf{\emph B}$. The velocity vector field ($\textbf{\emph V}$) is derived by using the Differential Affine Velocity Estimator for Vector Magnetograms (DAVE4VM) method \citep{sch08} in our study. The DAVE4VM method implements a variational principle with the magnetic induction equation to track the velocity of magnetic footpoints. We adopt 19 $\times$ 19 pixels as the window size for DAVE4VM, which is determined by examining the slope, Pearson linear correlation coefficient, and Spearman rank order between $\bigtriangledown_n \cdot (\textbf{V}B_n) $ and $\Delta B_n/ \Delta t$ \citep{sch06,sch08}. The first term in the right of the equation is named emergence term ($\dot H_e$), which is associated with the emergence of magnetic flux tube from the solar subsurface. The second term is named shear term ($\dot H_s$), which is generated by shearing magnetic field lines due to tangential motions on the surface \citep{ber84,kus02,par05,liu14}. As  the target active region is small enough, the curvature effect could be ignored. The solar photosphere S can be assumed to be a planar. Therefore, the helicity ejection flux can be expressed as follows \citep{par05,liu12}:
\begin{eqnarray}
\dot H &=& \dot H_e +\dot H_s \nonumber \\
       &=&\frac{1}{2\pi}\int_s \int_{s\arcmin}\textbf{\emph n}\cdot \frac{\textbf{\emph{x}}-\textbf{\emph{x}}\arcmin}{|\textbf{\emph{x}}-\textbf{\emph{x}}{\arcmin}|^2} \times \{\textbf{\emph B}_t(\textbf{\emph{x}})V_{\bot n}(\textbf{\emph{x}})B_n(\textbf{\emph{x}}\arcmin) \nonumber \\
       & & -\textbf{\emph B}_t (\textbf{\emph{x}} \arcmin)V_{\bot n}(\textbf{\emph{x}} \arcmin)B_n(\textbf{\emph{x}})\} dS\arcmin dS \nonumber \\
       & &-\frac{1}{2\pi}\int _s\int _{s\arcmin}\textbf{\emph n}\cdot \frac{\textbf{\emph{x}}-\textbf{\emph{x}}\arcmin}{|\textbf{\emph{x}}-\textbf{\emph{x}}{\arcmin}|^2} \times \{[\textbf{\emph V}_{\bot t}(\textbf{\emph{x}}) \nonumber \\
      & &-\textbf{\emph V}_{\bot t}(\textbf{\emph{x}}\arcmin)]B_n(\textbf{\emph{x}})B_n(\textbf{\emph{x}}\arcmin)\}dS\arcmin dS \nonumber \\
      &=& \int_s G_\theta(\textbf{\emph{x}}) dS,\label{eq13}
\end{eqnarray}
in which $\textbf{\emph{x}}$ and $\textbf{\emph{x}}\arcmin$ represent two photospheric positions and $\textbf{\emph n}$ is the surface normal vector pointing into the corona. $G_\theta (\textbf{\emph{x}})$ could be used to present the approximate proxy of the magnetic helicity injection flux density \citep{par05}.  We use Eq.\ref{eq13} to calculate the total helicity flux in the whole active region. Accumulated helicity can be derived by the integration of the helicity injection rate with time.

To minimize the influence of measurement error in the magnetic field of HMI, we only consider pixels with a magnetic field strength exceeding 300 G (roughly 3$\sigma$ noise level of the vector magnetic field) to calculated above magnetic parameters \citep{hoe14}. The errors of positive and negative fluxes are estimated by the uncertainties from HMI observations of SHARP data. On the other hand, a Monte-Carlo experiment was used to estimate the errors in Eqs.\ref{eq1}, \ref{eq2}, \ref{eq3}, \ref{eq7}, \ref{eq8}, \ref{eq11}, and \ref{eq13}. In the Monte Carlo experiment, we randomly added noises to three components of the vector magnetic field, and repeated the computations of above equations. The added noises have a Gaussian distribution, and the widths ($\sigma$) of Gaussian distributions are set to be the uncertainties from HMI observations in each components of the vector magnetic field. This test was repeated 200 times. The two-hour running average of the root mean square (rms; $\sigma$) of these 200 experiments is used to be the representative error as the method estimated by \cite{liu12}.
\section{Results} \label{sec:results}
\subsection{Emerging process of the active region}
\begin{figure*}
   \centering
   \includegraphics{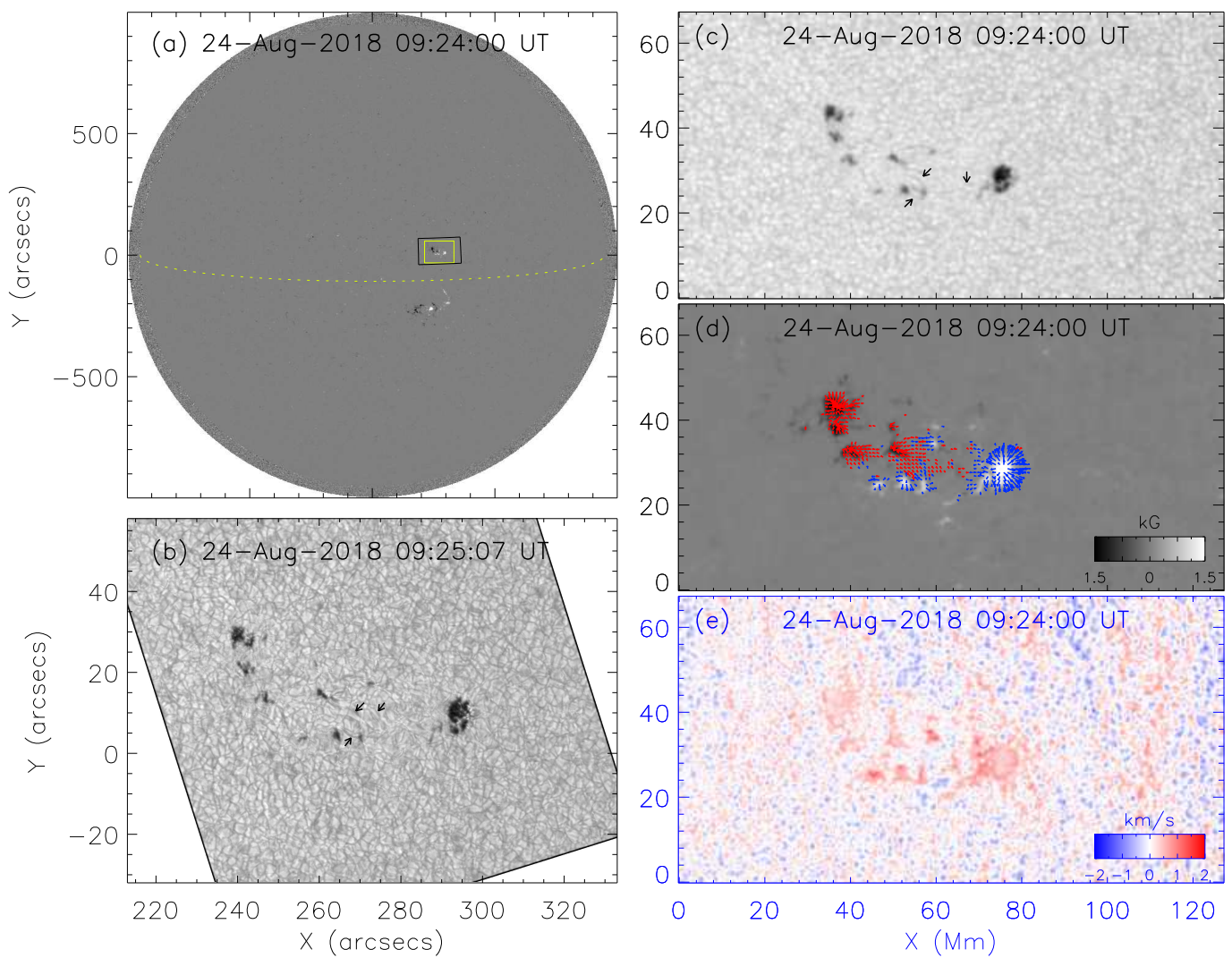}
   \caption{Overview of the emerging active region. (a) The global line-of-sight magnetic field observed by SDO/HMI. The yellow dashed line marks the solar equator. The yellow and black boxes outline the fields of view of panel (b) and panels (c)-(e), respectively; (b) TiO image observed by NVST; (c)  continuum intensity image from SHARP data; (d) vector magnetograms from SHARP data;  blue and red arrows indicate the transverse field with positive and negative flux, respectively; (e) Doppler velocity from SHARP data.}\label{fig1}
    \end{figure*}

During the period from August 24 to 25 2018, an active region NOAA 12720 gradually emerged into the solar surface in the northern hemisphere, and appeared near the center of the solar disk (N07, W16). Figure \ref{fig1} shows the overview of this emerging active region at around 09:24 UT on August 24, 2018. Panel (a) displays the full disk line-of-sight magnetic field observed by SDO/HMI. This active region occurred in a purely clean environment, which is a perfect isolated AR. Panel (b) shows the TiO image observed by NVST. A lot of elongated granules and stretched dark lanes marked by black arrows in panels (b) could be identified between two main pores or sunspots, which are considered to be the signatures of the emerging flux through the photosphere \citep{str96,sch10}. The panels of the right column exhibit the images from SHARP data at the corresponding moment. Panel (c) is the continuum intensity image. The elongated granules and stretched dark lanes marked by the black arrows also could be found in the flux-emerging region. Panel (d) shows the vector magnetogram at 09:24 UT on August 24. The horizontal magnetic field ($\textbf{B}_t$) is depicted by blue and red arrows indicating the direction and strength, while the background is the map of the vertical magnetic field component ($B_z$). The relatively concentrated leading positive polarity and the dispersed following negative polarity indicate the fact that this active region does not obey with the hemisphere pattern of Hale's law \citep{hal25} for the active region in northern hemisphere in the solar cycle 24. Thus, the emerging active region is an anti-Hale one. Panel (e) shows the Doppler velocity map in the photosphere. Blue and red colors indicate upflows and downflows, respectively. The fascinating feature is that the regions with strong magnetic flux show strong red shift in the photosphere. This may be associated with the falling plasma. As the magnetic fields ascend into higher altitude, the plasma residing in the emerging magnetic field would fall down through the footpoints along the magnetic field lines due to the solar gravity, and returns to the solar surface. These falling plasma shows the red shifted signatures at the footpoints of the magnetic field lines.

\begin{figure*}
\centering
\includegraphics{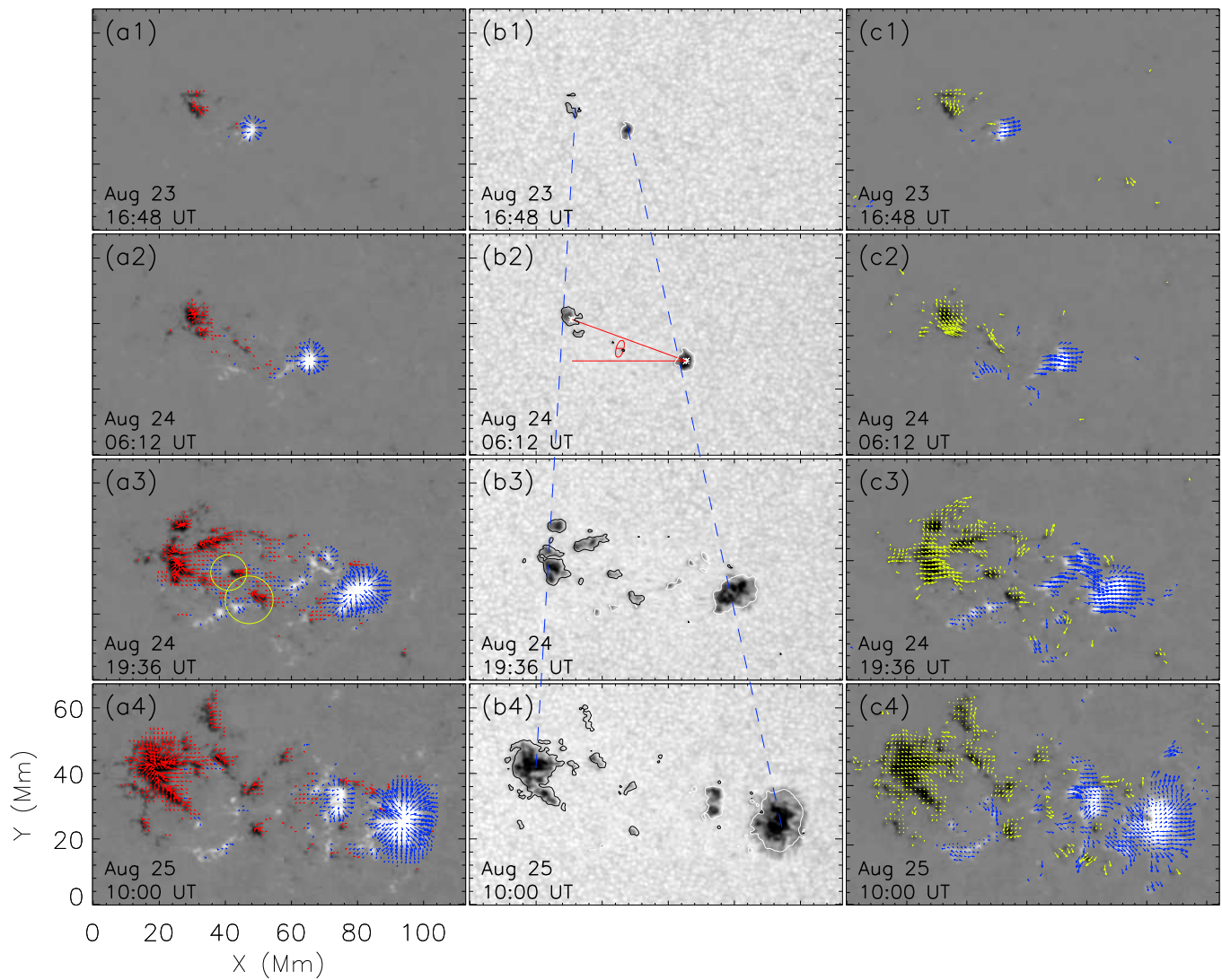}
\caption{Evolution of the emerging active region NOAA 12740. (a1)-(a4) The vector magnetograms. (b1)-(b4) The continuum intensity images. The white and black contours indicate the magnetic field strength (800G) with positive and negative polarities, respectively. The white asterisks in panel (b2) denote the centroids of positive and negative polarities; (c1)-(c2) The transverse velocity maps derived with DAVE4VM. The blue and green arrows indicate the transverse velocity with positive and negative polarities, respectively.}\label{fig2}
\end{figure*}

Figure \ref{fig2} exhibits the emerging process of the active region NOAA 12720. The panels of left column show vector magnetograms at different moments, while the ones in the mid and right columns are corresponding continuum intensity and horizontal velocity, respectively. The active region initially appeared as a simple small bipolar pore (see panels (a1) \& (b1)), which eventually developed into a large active region including a leading positive sunspot, a following negative sunspot and some mixed polarities between two main sunspots (see panels (a4) \& (b4)). The detailed evolution of the active region is displayed by the animation of the Fig. \ref{fig2}. Most of the newly emerging flux initially appeared between two main polarities and gradually accumulated toward each polarity. Some moving dipolar features \citep{ber02} could be found between two main polarities, which is marked by the yellow circle in the panel (a3). These types of structures are important to form the long coronal loop connecting two main polarities and release the mass from rising magnetic field at the later stage of emergence \citep{cen12}. Panels (b1)-(b4) show the evolution of two main sunspots during the emergence of the active region in continuum intensity images. The leading sunspot tends to be more concentrated than the following one. On the other hand, the leading sunspot drifted westward while the following sunspot drifted eastward. The leading sunspot moved faster than the following one, which is demonstrated by the inclinations of two dashed blue lines in the mid column. This feature is reasonable for the leading (or following) sunspot during emergence. In general, the leading sunspot is close to the solar equator and drift westward while the following sunspot is far away from the solar equator and drift opposite direction. With the influence of solar rotation and differential rotation in latitude, the leading sunspot is expected to drift faster than the following one during their emergence. Based on the velocity map derived by DAVE4VM in the panels of the right column, the positive polarity exhibited a strong motion toward the southwest at the early phase of emergence, while the negative polarity moved toward the northeast (panel (c1)-(c3)). At the later phase of emergence, these motions became weak and inconsistent (panel (c4)).

\begin{figure*}
\centering
\includegraphics{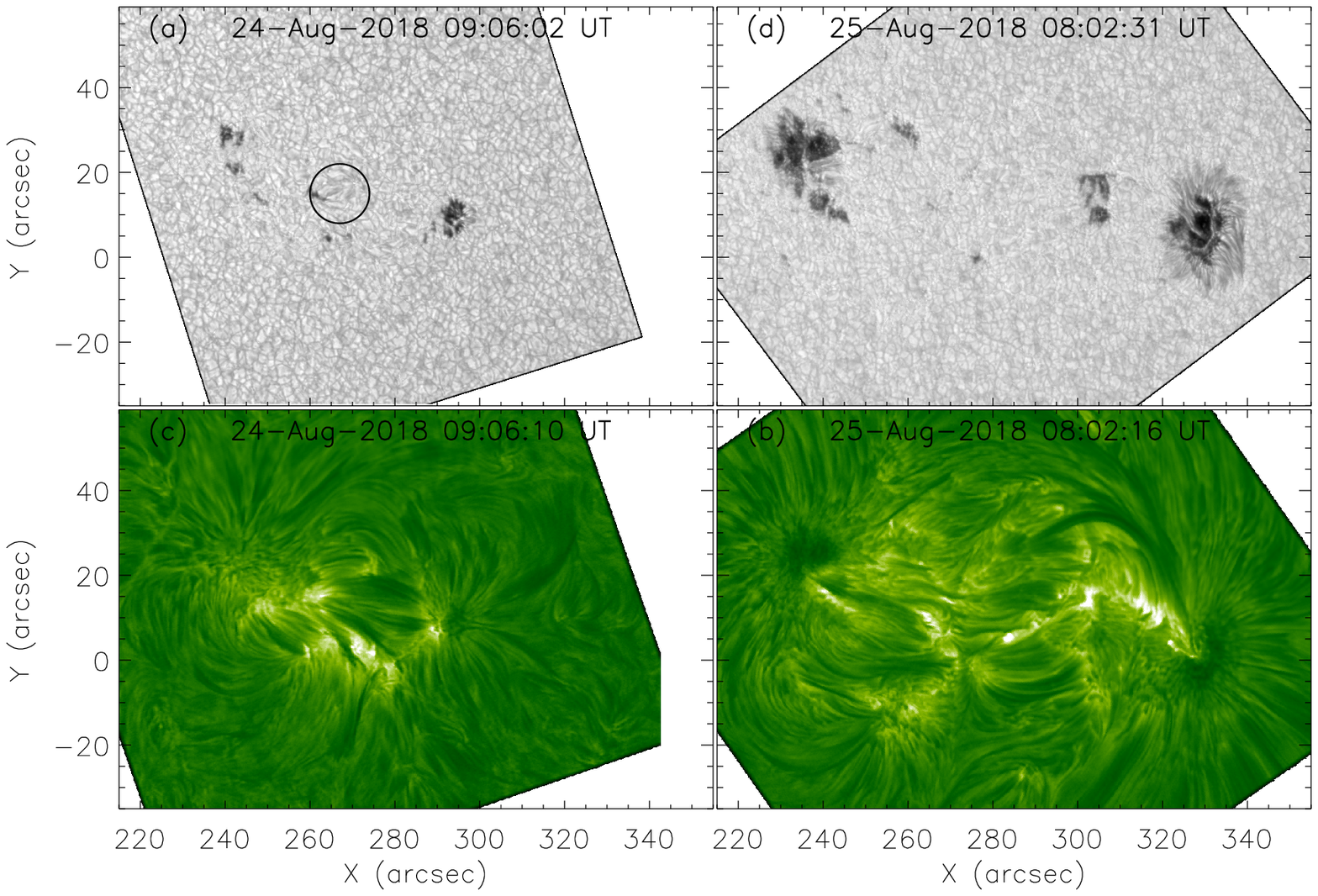}
\caption{Photospheric and chromospheric observations during the emergence. (a) \& (b) TiO images observed by NVST; (c) \& (d) H$\alpha$ images observed by NVST. Black circle in panel (a) points out some elongated granules and stretched dark lanes.}\label{fig31}
\end{figure*}

Figure \ref{fig31} shows the changes in photosphere and chromosphere between the early and latter stages of emergence. The images in top row are TiO observations from NVST, while the ones in bottom row are corresponding H$\alpha$ observations from NVST. At early stage of emergence, a lot of elongated granules and stretched dark lanes can be found in the middle of active region in the photosphere, which are marked by black circle in panel (a) of Fig. \ref{fig31}. Meanwhile, the sheared-arcade structures appeared in the chromosphere (see in panel (c)). At the latter stage of emergence, the active region was approaching mature, which consists of leading, following sunspots and some fragments. It is interesting to find that the leading sunspot has a complete penumbra while the following sunspot does not (see panel (b)). Meanwhile, more complicated structures (such as filaments or small-scale filamentary structures) could be seen in the chromosphere (see panel (d)).
\subsection{Magnetic flux, pole separation and tilt angle}
\begin{figure*}
\centering
\includegraphics{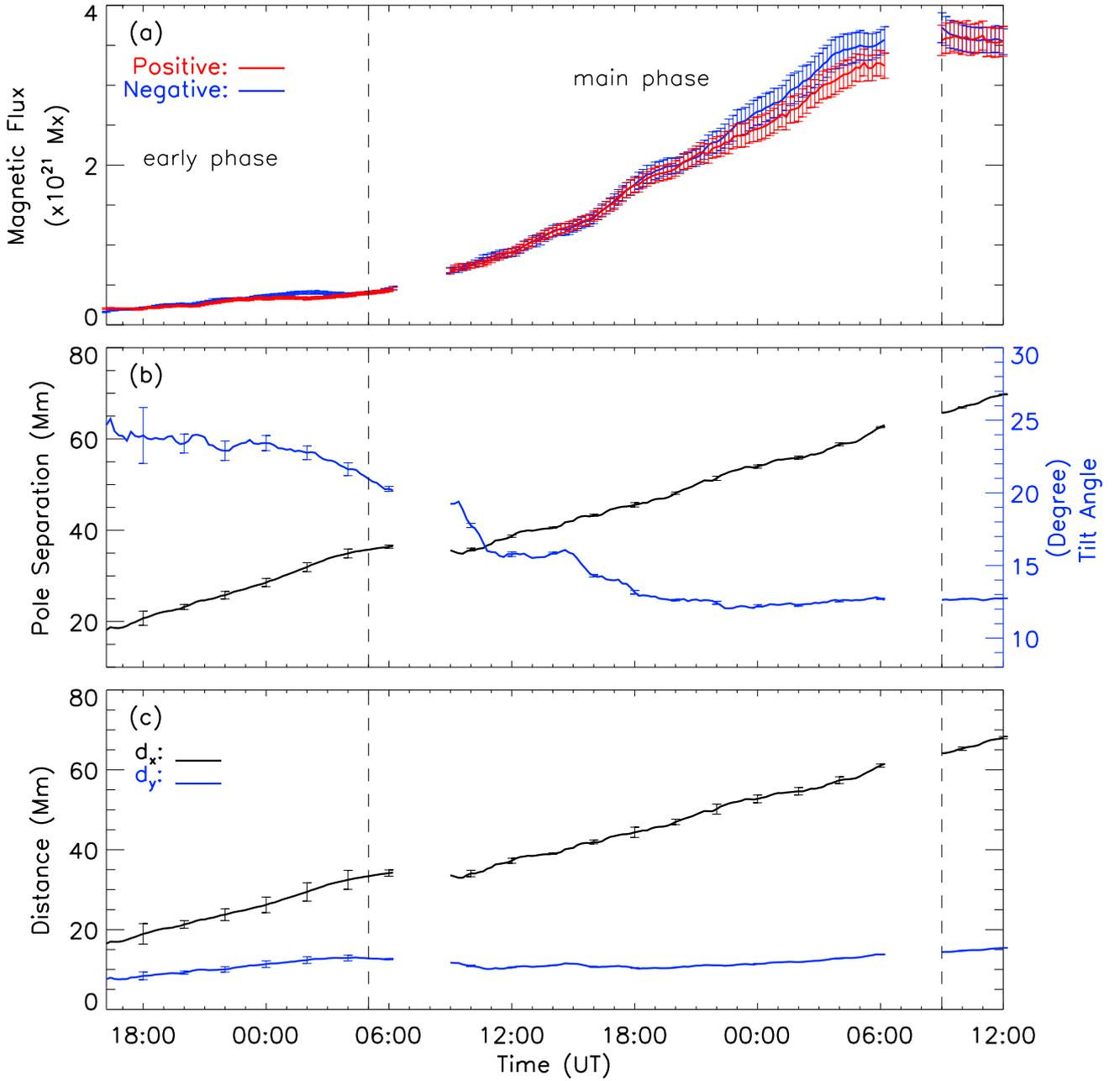}
\caption{Variations of magnetic flux, pole separation, and tilt angle of the emerging active region during the period from 16:12 UT on August 23 to 12:00 UT on August 25, 2018. (a) The evolution of positive and negative magnetic flux. The red and blue lines indicate the positive and negative magnetic flux. The errors in the positive and negative magnetic flux are estimated by the uncertainties from HMI observations of SHARP data; (b) The evolutions of pole separation and tilt angle. The black line denotes pole separation and the blue one denotes tilt angle; (c) The evolution of pole separation in longitude direction ($d_x$) and latitude direction ($d_y$). The black and blue lines denote the pole separation in longitude direction and latitude direction, respectively. Error bars, reported at several representative times in panels (b) \& (c), are estimated by using the Monte-Carlo method.}\label{fig3}
\end{figure*}

Based on the equations described in Sect. \ref{methods}, we calculate the magnetic flux, pole separation, and tilt angle of the emerging active region during its emergence. Figure \ref{fig3} (a) exhibits the temporal variations of positive and negative magnetic fluxes. The error bars are derived by integrating the uncertainties from HMI observation with integrated area. Both positive and negative magnetic fluxes show a equivalent increase from about 0.2 $\times 10^{21}$ Mx at 16:12 UT on August 23 to 3.6 $\times 10^{21}$ Mx at 12:00 UT on August 25. The mean increasing rate of magnetic flux is about 0.78 $\times 10^{20}$ Mx/hr in each polarity. We can find that both positive and negative magnetic fluxes maintained a slow increase during the period before 05:00 UT on August 24. The mean increasing rate of each polarity was only about 0.17 $\times 10^{20}$ Mx/hr during this period. We define this period as the early phase of emergence. Both positive and negative magnetic fluxes experienced a rapid increase during the period from 05:00 UT on August 24 to 9:00 UT on August 25. The mean increasing rate of each polarity was 1.16 $\times 10^{20}$ Mx/hr. We define this period as the main phase of emergence. Therefore, the emergence of this naked anti-Hale AR is also consistent with the two-step emergence pattern \citep{fu16}. Unfortunately, the data during the periods from 06:24 to 08:48 UT on August 24 and 25 were unavailable because of the eclipse in the SDO orbit \citep{pes12}. Both positive and negative fluxes exhibit little change during the period from 09:00 UT to 12:00 UT on August 25. This indicates that the active region became mature at around 12:00 on August 25. Panel (b) shows the time variations of tilt angle and pole separation. The blue line denotes the tilt angle of the active region, while the black line denotes the pole separation. The uncertainties are depicted by the black and blue representative error bars. They are estimated by conducting a Monte Carlo experiment, as described in Sect. \ref{methods}. There are several features on the profile of tilt angle. Firstly, at the early phase of the flux emergence, the active region maintained a high value for the tilt angle (about 23$\degr$) and began to decrease. Secondly, the main decrease of the tilt angle took place during the phase from 03:36 UT to 18:00 UT on August 24. Thirdly, the active region kept at low value of tilt angle (about 12$\degr$) after 18:00 UT, when the magnetic flux continued to increase. The pole separation shows a continuously increasing pattern with a constant rate during the entire time, which was from about 18 Mm at the beginning to about 70 Mm in the end. The mean increasing rate of pole separation was about 2.92 Mm/hr. We divide the pole separation into two components: longitude component, $d_x$, and latitude component, $d_y$. Panel (c) shows the evolutions of $d_x$ and $d_y$. The longitude component $d_x$ increased significantly all the time from about 16 Mm to 68 Mm. The enhancement of pole separation in latitude direction ($d_y$) is much smaller than that in the longitude direction. Therefore, the increase of the polar separation is mainly contributed from the enhancement in the longitude direction ($d_x$). This means that the two poles are separated from each other in longitude as the active region emerges.

\begin{figure*}
\centering
\includegraphics{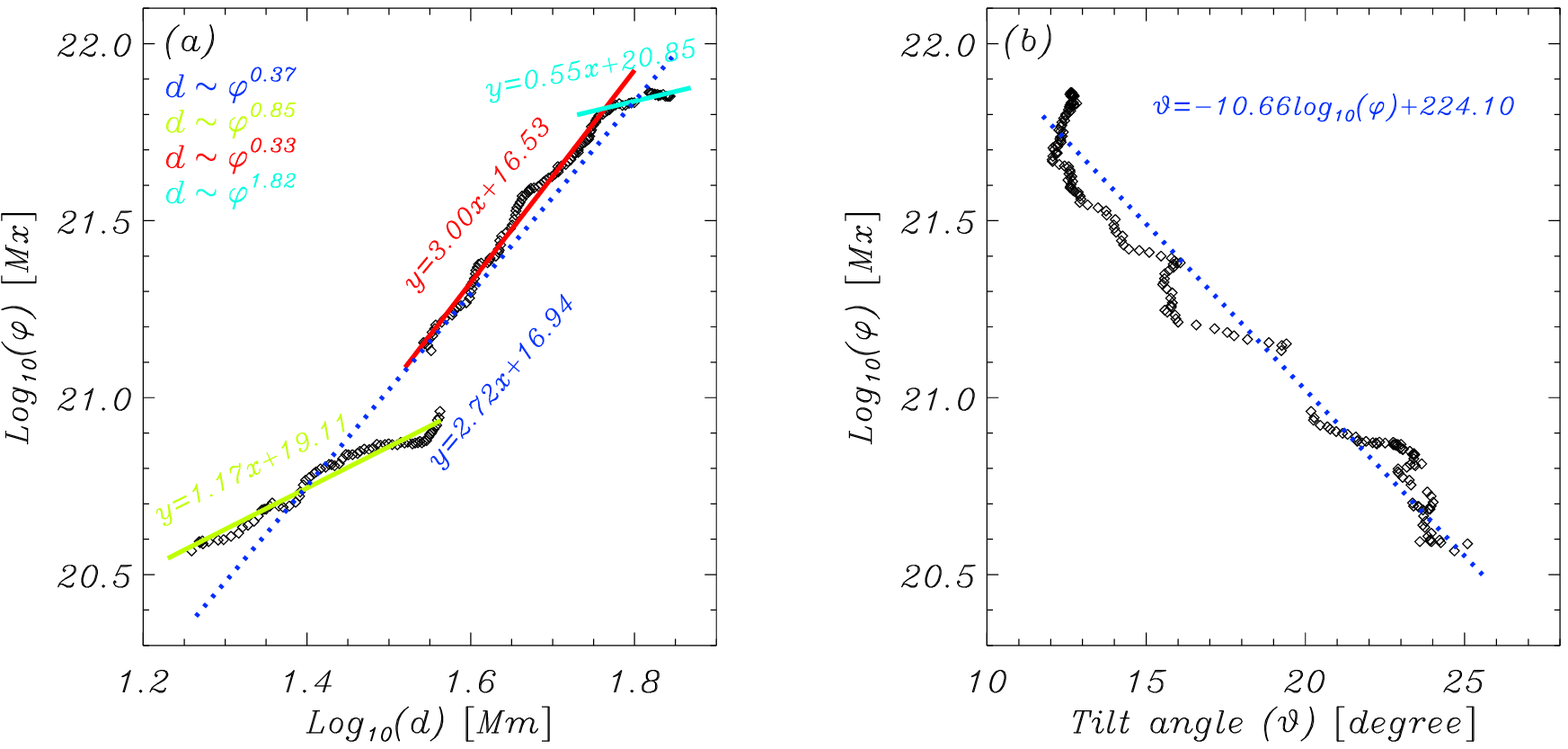}
\caption{Correlation between the magnetic flux and pole separation or tilt angle; (a) profile of total magnetic flux as a function of the pole separation. The different color lines denote the fitting lines with different slopes. The different color functions depict the results of different fitting lines by using the linear fit; (b) profile of total magnetic flux as a function of the tilt angle. The fitting line by linear function is indicated by the blue dotted line.}\label{fig4}
\end{figure*}

In order to investigate the relationships between the magnetic flux and pole separation or tilt angle during the emergence of the active region, we plot the total magnetic flux as a function of pole separation and tilt angle in Fig. \ref{fig4}. Figure \ref{fig4} (a) shows the total magnetic flux ($\varphi$) as a function of the pole separation ($d$). A positive correlation between these two parameters is found. By using a linear fitting for the logarithm of these variables, a correlation of $log_{10}(\varphi)=2.72*log_{10}(d)+16.94$ plotted by blue dotted line can be derived. This correlation is equivalent to $d$ $\sim$ $\varphi^{0.37}$. In detail, when the pole separation is smaller than $\sim$ 36.5 Mm ($log_{10} d$ $<$ $10^{1.56}$ Mm), we get that  $ log_{10}(\varphi)=1.17*log_{10}(d)+19.11$ shown by the yellow line and is equivalent to $d$ $\sim$ $\varphi^{0.85}$. When the pole separation is between 36.5 Mm and 58.4 Mm ($10^{1.56}<log_{10}d$ $\leqslant10^{1.77}$ Mm), the logarithm of magnetic flux equals to 3.00$*log_{10}(d)+16.53$ shown by the red line, which is equivalent to $d$ $\sim$ $\varphi^{0.33}$. When the pole separation is larger than 58.4 Mm ($log_{10}d$ $>$ $10^{1.77}$ Mm), the logarithm of magnetic flux equals to 0.55$*log_{10}(d)+20.85$ plotted by the pink line, which is equivalent to $d$ $\sim$ $\varphi^{1.82}$. In other words, during the emergence of active region, the magnetic flux has a relatively small increasing rate when the pole separation is shorter than 36.5 Mm and longer than 58.4 Mm. The magnetic flux experiences a significant increase at the moderate pole separation of the range from 36.5 Mm to 58.4 Mm. This fact manifests that the magnetic flux of naked AR main increases in the moderate pole separation. Fig. \ref{fig4} (b) exhibits the total magnetic flux as a function of the tilt angle. We also use a linear function to fit the scatter plot. A correlation of $\vartheta$ $=$ $-10.66*log_{10}(\varphi)+224.10$ can be derived. This means that there is a negative correlation between the tilt angle and magnetic flux during the dipolar AR emergence. In detail, the active region with small magnetic flux ($<$ 10$^{20.8}$ Mx) kept a high and almost constant tilt angle of $\sim$ 23$\degr$-24$\degr$. The active region with large magnetic flux ($>$ 10$^{21.6}$ Mx) kept in a small and almost constant tilt of $\sim$ 12$\degr$-13$\degr$. The tilt angle decreased gradually from about 23$\degr$ to 13$\degr$, when the total magnetic flux increased from 10$^{20.8}$ to 10$^{21.6}$ Mx. This also manifests that the newly emerging magnetic flux with a high tilt angle emerges from subsurface at first. And then two polarities of emerging magnetic flux separate mainly from each other in longitude direction, which results in the decrease in tilt angle.
\subsection{Magnetic properties}
\begin{figure*}
\centering
\includegraphics{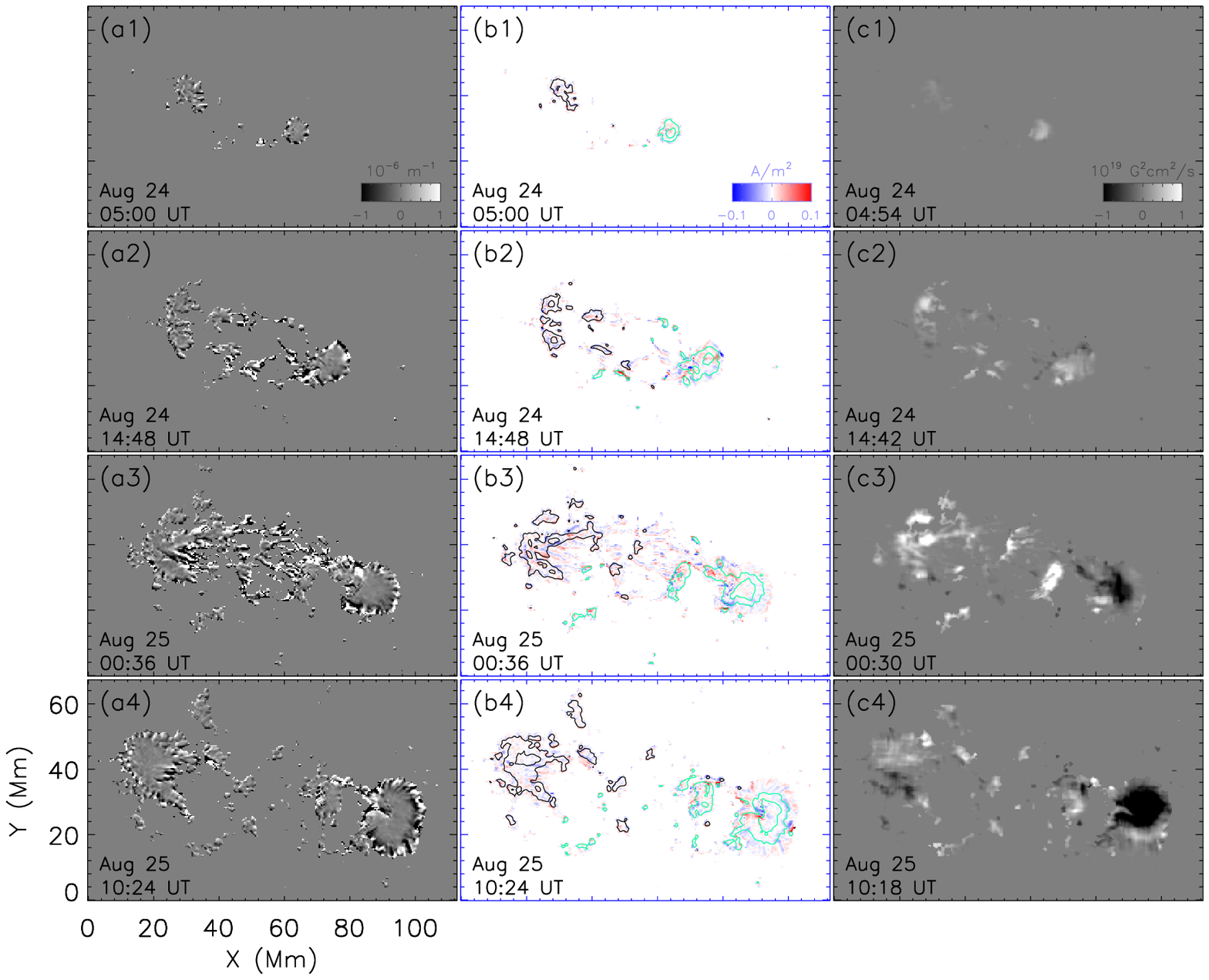}
\caption{Distributions of force-free parameter ($\alpha_z$), vertical electric current density, proxy of heilicty flux density ($G_\theta$); (a1)-(a4) the force-free parameter ($\alpha_z$); (b1)-(b4) the vertical electric current density ($J_z$). The pink and black contours denote the positive and negative magnetic field with the levels of 500 G, 1500 G, respectively; (c1)-(c4) the proxy of helicity flux density (G$_\theta$). It only plots the pixels with the total magnetic field strength bigger than 300 G.}\label{fig5}
\end{figure*}

Figure \ref{fig5} shows the distributions of the force-free parameter $\alpha_z$, vertical electric current ($J_z$), and proxy of heilicty flux density ($G_\theta$) in the active region. Panels (a1)-(a4) exhibit the distributions of the force-free parameter $\alpha_z$ in the photosphere at different moments. The force-free parameter shows a disordered pattern in each polarity. It is interesting to find that the strong force-free parameter mainly located in the periphery of main polarities. Panels (b1)-(b4) show the distributions of the vertical electric current density ($J_z$) in the photosphere at the corresponding moments. We also find that the vertical electric current shows a disordered in each polarity and the strong vertical electric current also located in the periphery of main polarities. Panels (c1)-(c4) display the distributions of magnetic helicity ejection flux density ($G_\theta$) in the photosphere at different moments. The density of the helicity ejection flux shows a relative uniform pattern unlike the force-free parameter and vertical electric current (positive and negative helicity ejection fluxes are dominant in each polarity). At around  04:54 UT and 14:42 UT on August 24, the helicity ejection flux was small and dominated by positive value (see panels (c1)-(c2)). At around 00:30 UT on August 25, it was also dominated by positive helicity ejection flux for the whole active region. However, the leading sunspot (positive polarity) was dominated by negative helicity ejection flux (see panel (c3)). At around 10:18 UT on August 25, the negative helicity ejection flux was dominant in the active region, especially in the leading sunspot (positive polarity). 

\begin{figure*}
\centering
\includegraphics{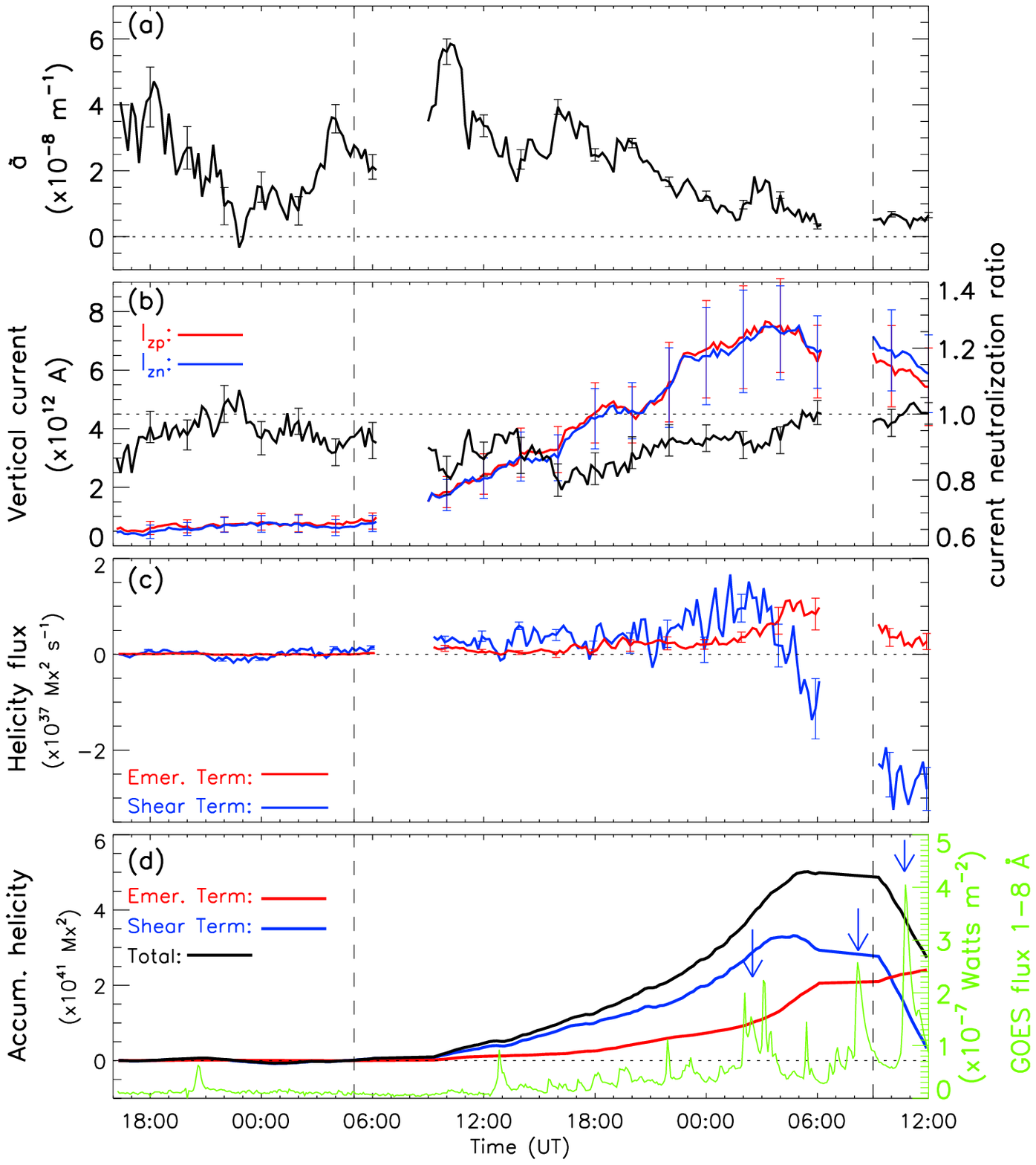}
\caption{Evolution of different magnetic parameters in the active region during the period from 16:12 UT on August 23 to 12:00 UT on August 25, 2018: (a) The flux-weighted force-free parameter ($\tilde{\alpha}$); (b) Vertical electric currents and the current neutralization ratio. Red and blue lines denote the positive and negative vertical current, respectively. The black line plots the profile of current neutralization ratio; (c) The magnetic helicity injection flux. Red and blue lines denote the emergence term and shear term, respectively; (d) Accumulative helicity. Red and blue lines indicate the accumulative helicity contributed from emergence term and shear term. The total accumulative helicity is shown by the black line. The green line denotes the  flux of GOES soft X-ray 1-8 $\rm\AA$. The two vertical dashed lines denote the times of 05:00 UT on August 24 and 09:00 UT on August 25. Some B-class flares are marked by blue arrows. Error bars estimated by the Monte-Carlo experiment are plotted at representative times.}\label{fig6}
\end{figure*}

Figure \ref{fig6} shows the evolution of different magnetic parameters in the whole active region during the emergence of the active region. Panel (a) exhibits the variation of the average flux-weighted force-free parameter $\tilde{\alpha}$ calculated by the Eq.\ref{eq11}. The average force-free parameter is almost positive during the emergence of the active region, which does not follow the hemisphere rule. The magnitude of the average force-free parameter is the order of $\sim$ $10^{-8}$ m$^{-1}$, which is consistent with results from previous studies \citep{wan00,liu14,kut19}. It rises up to 6 $\times$ $10^{-8}$ m$^{-1}$ at 10:00 UT on Aug 24. At the end of the emergence, the average force-free parameter is only $\sim$ 0.5 $\times$ $10^{-8}$ m$^{-1}$. In contrast to the result from \cite{wan00}, it is variant instead of constant during the emergence of the active region. Panel (b) exhibits the evolutions of the vertical electric current and the current neutralization ratio ($\Re$). The red line indicates variation of the positive electric current while the blue line indicates variation of the negative electric current. The positive and negative electric currents are almost equivalent during the entire emergence of the AR. This fact manifests that no any net current is carried into the atmosphere during the emergence of this isolated AR. It is reasonable to assume that the total current in the isolated AR stays neutralized during its emergence; this is because the total current $I_{total}$ over the whole photospheric AR is equal to $\frac{c}{4\pi}\oint \bf{B_t} \cdot \it d \bf{l,}$ and  $\bf{B_t}$ would be balanced along the path outside the  boundary of the isolated AR \citep{par96}. At the early phase of emergence, they remained in low magnitude of less than 1 $\times$ $10^{12}$ A. At the main phase of emergence, they experienced a rapid increase at first and then decreased at the latter phase. During the period from around 03:48 UT on Aug 24 to 03:24 UT on Aug 25, both positive and negative electric currents increased mainly from about 0.5 $\times$ $10^{12}$ A to about 7.5 $\times$ $10^{12}$ A. After around 04:00 UT on Aug 25, they experienced a decrease until the end time of investigation. The peak time of electric current (about 03:24 UT on August 25) was earlier than that of the magnetic flux (about 10:00 UT on August 25) (see Figs. \ref{fig3} \& \ref{fig6}). The black line shows the variation of the current neutralization ratio ($\Re$). The $\Re$ values are in the range from about 0.8 to 1.1, which suggests that the AR kept neutral current during its emergence. This result is consistent with the previous observational studies, which also supports the premise that currents become more neutralized during magnetic flux emergence in flare-quiet ARs \citep{liu17,ava20}.

With the methods described in Sect. \ref{methods}, we also calculated the magnetic helicity injection flux of the active region during the emergence of the active region. According to Eq.\ref{eq13}, we could obtain the magnetic helicity injection flux from the emergence term and the shear term. Figure \ref{fig6} (c) exhibits the evolution of magnetic helicity injection flux. The magnetic helicity flux contributed by the emergence term is plotted by the red line while the magnetic helicity flux contributed by the shear term is plotted by the blue line. There are several features in the magnetic helicity flux. Firstly, both terms of the helicity injection flux kept a low value at the early phase of the emergence. Secondly, the shear term changed its sign from positive to negative at about 04:00 UT on Aug 25. Thirdly, the emergence term stayed in positive during the whole period. On the other hand, we calculate the accumulated helicity by the time integral of the helicity injection flux with setting the accumulated helicity to be zero at the beginning of investigation. Figure \ref{fig6} (d) shows the time variation of accumulative helicity. At the early phase, the accumulative helicity was very low. At the main phase of the emergence, the total accumulative helicity also showed the same pattern as the vertical current, which was increasing at first and then began to decrease. Overall, this active region is mainly dominated by the positive helicity. This feature does not obey the hemisphere rule (being positive in the northern hemisphere and negative in the southern hemisphere at the 24 solar cycle). Furthermore, during the most of time, the total magnetic helicity in the active region was contributed mainly by the shear term, which is also consistent with previous findings \citep{liu12,liu14,wan19}.

The green line in panel (d) denotes the variation of soft X-ray 1-8 $\rm\AA$ flux during the emergence of the active region. We can find that several B-class flares took place in this active region. These flares marked by blue arrows in panel (d) mainly appeared at the latter phase of the emergence. As is well known, both the electric current and magnetic helicity are the proxies of non-potential magnetic field. In general, the flares would transport the magnetic energy to other forms of energy (such as thermal energy and kinetic energy of plasma) by magnetic reconnection. It implies that these flares may be associated with the decrease of electric current and the magnetic helicity at the latter phase.

\subsection{Magnetic twist}
\begin{figure*}
\centering
\includegraphics{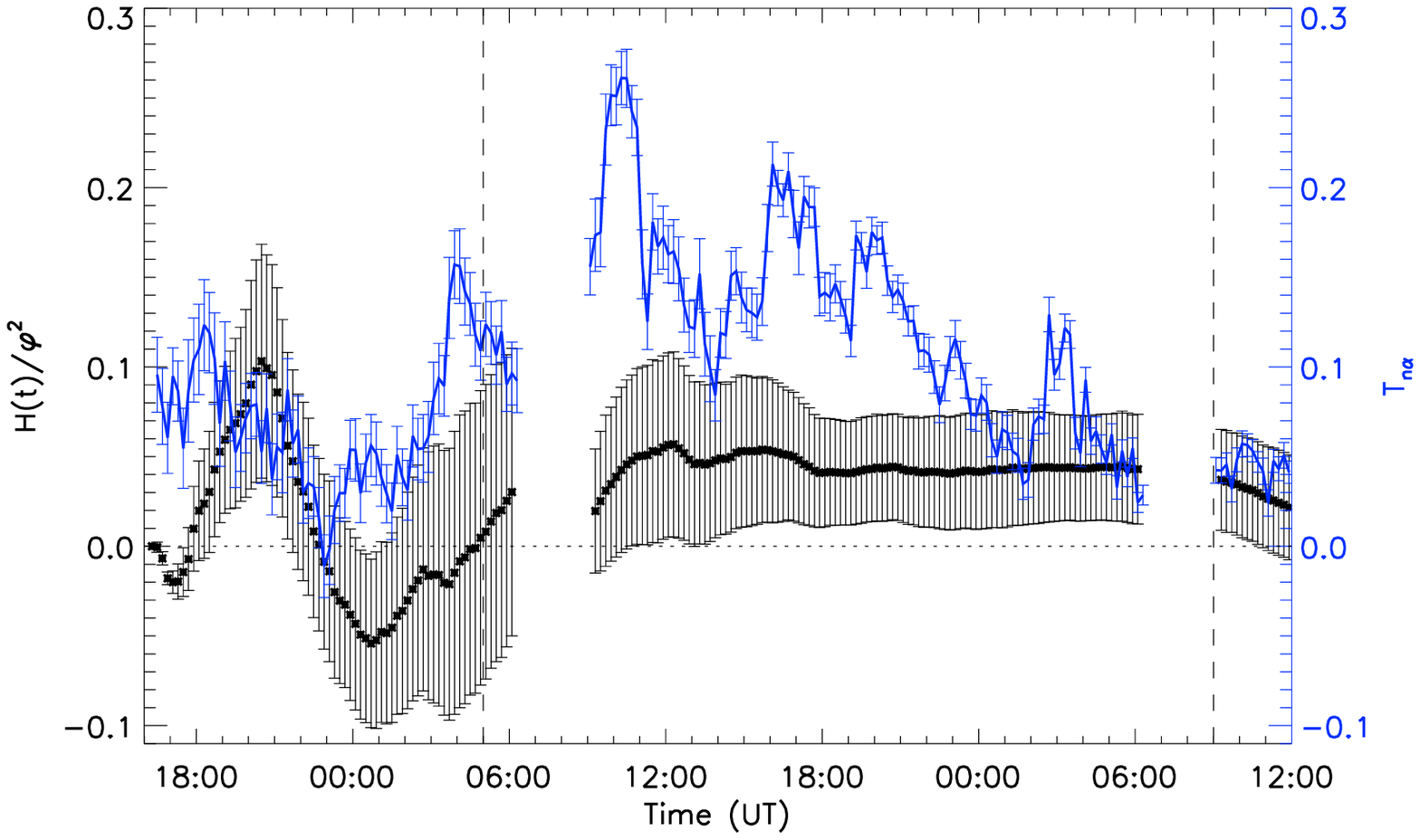}
\caption{Evolution of twist numbers ($\bar{n}$ and $T_{n\alpha}$) in the entire active region during the emergence of the active region. The asterisk denotes the profile of $\bar{n}$ while the solid line denotes the one of $T_{n\alpha}$. The horizontal dotted line denotes zero of twist number. The two vertical dashed lines denote the times of 05:00 UT on August 24 and 09:00 UT on August 25. The errors are transferred from the parameter's errors estimated by the Monte-Carlo experiment. }\label{fig7}
\end{figure*}

On one hand, the magnetic helicity inside a uniformly twisted flux tube with n turns, can be estimated by $\rm H = n \varphi^2$ \citep{ber84}, where $\varphi$ is axial magnetic flux. Therefore, if we assume that the magnetic field of active region is a uniformly twisted flux tube, the twist number of active region can be derived by the magnetic helicity and magnetic flux. The black  line of asteriskes in Figure \ref{fig7} plots the twist number, $\bar{n} \approx \rm H(t)/\bar{\varphi}^2$, where $\bar{\varphi}$ is the average magnetic flux between two polarities ($\bar{\varphi}=(\varphi_{zp}+\varphi_{zn})/2$) and H(t) is total accumulated helicity. This quantity ($\bar{n}$) indicates how much twist is in the the magnetic configuration of the entire active region. At the early phase, the twist number had changed its sign several times. This feature may be related to the small and mixed helicity injection flux from the subsurface at the early phase. At the main phase, the twist number increased to about 0.05 at first and then it lowered slightly and remained at about 0.04. After that, it experienced a sharp decrease until the end of our investigation time frame. Except for the early phase, the twist number almost kept in positive during the emergence of the active region. Due to the small magnetic helicity and magnetic flux at the early phase, we only focus on the twist number in the main phase of emergence. We were able to find that the twist number was on the order of 10$^{-1}$ during the emergence. The positive sign of the twist number is consistent with the positive sign of force-free parameter, $\tilde{\alpha}$, which does not follow the hemisphere rule either. On the other hand, we assume that the emerged magnetic field is a linear force-free field. The force-free parameter is set to be $\tilde{\alpha}$. Therefore, the twist number (T$_{n\alpha}$) of magnetic field could be estimated by the function of T$_{n\alpha}=\frac{1}{4\pi}\int \tilde{\alpha}dL$, in which the L is the length of the magnetic field and simply assume to be $\frac{\pi d}{2}$. Based on the above equation and reasonable assumptions, the T$_{n\alpha}$ could be calculated by using the evolutions of $\tilde{\alpha}$ and pole separation (d). The profile of T$_{n\alpha}$ shows by the solid blue line as the function of time in Fig. \ref{fig7}. Interestingly, T$_{n\alpha}$ was also on the order of 10$^{-1}$ turns, which is consistent with the twist number ($\bar{n}$) derived by the helicity calculation. In detail, the T$_{n\alpha}$ is slightly bigger than $\bar{n}$ at the early stage of main emergence. At the end of emergence,  the T$_{n\alpha}$ is also about 0.04 turns, which is consistent with $\bar{n}$. This might be related to the continuous increasing twist in the emerged magnetic field at the early stage of main emergence. Nevertheless, both T$_{n\alpha}$ and $\bar{n}$ show almost the same value at the end of emergence. These results manifest that the twist of this naked anti-Hale active region is very low and in the order of 10$^{-1}$ turns. The errors of $\bar{n}$ mainly come from the uncertainty of accumulated helicity related to photospheric velocity and magnetic field, while the errors of T$_{n\alpha}$ mainly come from the uncertainty of force-free parameters related to magnetic field. We use a Monte Carlo experiment method to estimate the uncertainty of these parameters. Therefore, the errors of T$_{n\alpha}$ and $\bar{n}$ can be estimated by the propagation from these parameters' errors, which are shown by using error bars in Fig. \ref{fig7}. On the other hand, the lack of data during periods from 06:24 to 08:48 UT on August 24 and 25 might lead to an underestimation of the accumulated helicity, which results in the underestimation of $\bar{n}$. Furthermore, we use a photospheric averaged flux-weighted force-free parameter $\tilde{\alpha}$ to represent the force-free parameter of magnetic field, which might bring on some uncertainties in estimating T$_{n\alpha}$. Therefore, we take a conservative estimate that the twist number of this emerging active region is on the order of 10$^{-1}$ turns. Nevertheless, we use two different methods to estimate the twist number of the active region and obtain the consistent result, which demonstrates that the result is reliable.
\section{Summary and discussion}\label{sec:conclusion}
This paper presents a detailed study on the evolution and magnetic properties of an anti-Hale emerging active region NOAA 12720. Our main results are as follows.

(1) During the emergence of the anti-Hale active region, the leading sunspot with positive magnetic field drifted faster than the following one with negative magnetic field. Most of the newly emerging magnetic fluxes initially occurred between the two main polarities. 

(2) A positive correlation between magnetic flux and pole separation is found during the emergence. The increase of the pole separation is mainly contributed from the enhancement in the longitudinal direction. The tilt angle of the active region maintains a high value (about 23$\degr$) at the early phase and a low value (about 12$\degr$) at the latter phase. In contrast to the pole separation, the tilt angle shows a negative correlation with total magnetic flux. On the other hand, a power-law relationship between pole separation and total flux is found during the emergence of the active region. 

(3) The force-free parameter ($\alpha_z$) and vertical electric current density ($J_z$) exhibit a disorderly  distribution in each polarity. The average flux-weighted force-free parameter and accumulative helicity in the entire active region was positive, a result that does not obey the hemispheric rule. The electric current and magnetic helicity in the active region are mainly built up at the main phase of the emergence. In addition, the magnetic helicity injection in the active region is mainly contributed by the shear term.

(4) Through the calculation of the helicity and the force-free parameter to estimate the twist number, we infer that the twist number of the whole active region is on the order of 10$^{-1}$ turns.

According to the statistical result of 4385 sunspot groups determined by \cite{li18}, only 8.1 $\pm$ 0.4 \% of these active regions are the anti-Hale sunspot groups. Therefore, anti-Hale active region is relatively rare. This anti-Hale active region NOAA 12720 emerged in a clean environment, which provides a unique opportunity for us to investigate the nature of emerging magnetic flux and anti-Hale active region. We find that the active region emerged with a high tilt angle at the early phase. The tilt angle of the active region decreased gradually when the active region continued to emerge. The tilt angle of the active region came up to a low value at the mid-phase of emergence and kept invariant until the end of the emergence. If we hypothesize that the anti-Hale active action before emergence is a flux tube embedded in the subsurface, the evoution of the tilt angle in this anti-Hale active region is consistent with the emergence of the flux tube with low twist. As a flux tube emerges with low twist from the convective zone into the solar atmosphere, the tilt angle of two main polarities exhibits high value at the initial stage and then decreases to a lower value. As the bottom of the lifting flux tube escapes, the tilt angle remains invariant at a low value \citep{kni18}. A negative correlation between tilt angle and magnetic flux could be found during this bipolar AR emergence. However, \cite{kos08} studied 715 bipolar ARs and deduced that the tilt angle does not dependent on the amount of magnetic flux. These facts manifest the finding that the evolution of tilt angle depends on the magnetic flux during the emergence of one active region but the tilt angle does not depend on the amount of magnetic flux in various active regions.

During the emergence of this naked active region, the pole separation has a power-law dependence on the magnetic flux. The power-law index in our study is 0.37. Interestingly, \cite{kut19} also found a similar relationship between pole separation and maximum of magnetic flux using 423 ARs. Their power-law index is about 0.36 which is a similar value to ours. It should be noted that our study just focuses on the evolution of one active region, while the statistical study of \cite{kut19}  includes the properties of many different emerging active regions. Therefore, the dependence of polarities separation on magnetic flux is a fundamental relationship -- either in the evolution of one emerging active region or between different emerging active regions. On the other hand, there are three states on the relation between pole separation and magnetic flux (see Fig. \ref{fig4}(a)). When the pole separation is shorter than 36.5 Mm and longer than 58.4 Mm, the magnetic flux increases with a relatively small increasing rate. When the pole separation is moderate in the range from 36.5 Mm to 58.4 Mm, the magnetic flux increases with a large increasing rate. The similar relationship between pole separation and magnetic flux is also seen in the emergence phase of NOAA active region 8167 (see Fig. 1 (f) in \cite{kos08}). Combining  (a) \& (b) in Fig. 3, we see that these three states are mainly associated with the evolution of magnetic field while the pole separation increase with an almost constant rate. The first two states may be related to a two-step emergence pattern in the active region \citep{fu16}, with a relatively gradual emergence followed by a relatively sudden change to more rapid emergence. This feature has been seen in simulations of ``two-step'' model \citep{mat93,mag01,tor10}.  In models of two-step emergence,  the rate of emergence in the emerging flux tube becomes rapidly increasing from a relatively small as the emerging flux tube reaches around the photosphere or chromosphere. In addition, \cite{chi13} suggested that the emerging magnetic structure with a weaker-field structure preceding a stronger-field structure is responsible for the two-step emergence of AR 11158. The third presumes that the magnetic flux has a low increasing rate with pole separation at the later stage, which might be related to emergence of the serpentine magnetic field at the latter stage of emergence. These serpentine magnetic fields would release mass through magnetic reconnection \citep{che10,cen12}, resulting in the cancellation of magnetic flux on the photosphere. The cancellation of magnetic flux would slow down the growth of magnetic flux in the whole active region. Otherwise, these cancelled magnetic fluxes were often seen as the signatures of moving dipolar features (see Fig. \ref{fig2} (a3)) on the photosphere before cancellation.

The injection of magnetic helicity in this anti-Hale active region was contributed mainly from the shear term, which is consistent with previous results \citep{liu12,liu14}. According to the 3D numerical simulation of a twisted flux tube, \cite{mag03} found that relative magnetic helicity is dominated by the emergence term at the early stage of the emergence. However, we does not find this feature in our study. This may be related to the low twist in this emerging active region. We obtain that the twist number of the emerging active region is in the order of  10$^{-1}$ turns. This means that the emerging active region carried the low twist into the atmosphere. The low twist in the magnetic flux leads to a little injection of magnetic helicity contributed by the emergence term at the early stage. In other words, the low magnetic helicity injection flux contributed by the emergence term at the early stage manifests that the twist of the emerging magnetic flux is low.

According to the definition of the twist number in the axial symmetric cylindrical flux tube, we can get $B_\theta / B_l$ = $n*\frac{2\pi r}{L}$, where $B_\theta$ and $B_l$  are, respectively, the magnetic field strength of azimuthal and longitudinal components, while r is the radius of flux tube and L is the length of the flux tube. According to the definition of the twist parameter ($q=\frac{B_\theta}{rB_l}$), we can obtain that q$H_0$ = $n*\frac{2\pi H_0}{L}$, where $H_0$ is the pressure scale height ($H_0$=170 km) \citep{tor11}. With the assumption that the whole emerging active region is a flux tube, the length L equals $\frac{\pi d}{2}$. At 04:00 UT on August 25, the pole separation (d) of the active region equals to about 60 Mm. Combinated with the above values of the parameters, the twist parameter (q$H_0$) in this active region is estimated to be about 1 $\times$ 10$^{-3}$. Comparing with the critical value (q$H_0$ = 0.1) for the emergence of flux tube derived by \cite{tor11}, the twist of this emerging active region is relatively weak. Therefore, this result is opposed to the conclusion in which the flux tube with weak twist would fails to emerge into the upper atmosphere \citep{mur06,tor11}. 

While the magnetic twist plays an important role in the activity and evolution of the active region,  the studies of twist in the active region are needed to reveal the intrinsic properties of magnetic field \citep{nan06,luo11,poi15}. Our result indicates that the low twist in the sub-photospheric emerging part of a flux rope is coherent with previous studies \citep{liu06,poi15}. \cite{yan09} studied 58 emerging Hale active regions and obtained that the average twist number in these Hale active region is about 0.039 turns, a value that is comparable to our result. This implies that either Hale or anti-Hale active regions operate under the same rising mechanism in the solar interior. Otherwise, this can also demonstrate that this anti-Hale active region is likely to be generated by rising of toroidal flux tube with opposite orientation in the interior \citep{ste12}, instead of causing by the kink instability mechanism of highly twisted flux tube \citep{nan06,kni18}. However, the positive sign of twist including positive twist number and average force-free parameter ($\tilde{\alpha}$) is opposite the sign of structural chirality of solar active region in northern hemisphere \citep{see90}. The active region violating Hale rule might be responsible for this phenomenon.

Here, we consider the role played by the twist in the emergence of the magnetic flux tube. In our observational study, we use two different methods which are based on magnetic helicity calculation and force-free parameter to estimate the twist in this anti-Hale active region. We find that the emerged magnetic field covering the whole active region exhibits a lowly twisted structure. The twist number is in the order of 10$^{-1}$ turns. More recently, \cite{kni18} simulated the emergence of different twisted flux tubes from the convection zone into the corona. They found that the behaviours of low twisted flux tube emergence matched the observed properties of the single active region (such as $\beta$ style), while the ones of high twisted flux tube matched the complex active region ($\delta$ style). Moreover, some authors found that an untwisted flux tube could also be able to rise through the upper convection zone and emerge into the photosphere to form spots \citep{ste11,rem14,kni20}. \cite{kni20} explained that the rise of untwisted toroidal flux rope is triggered by the undular instability instead of buoyant instability. On the other hand, most studies of numerical simulation often set a mature magnetic flux rope embedding in the convection zone bottom as the initial condition. The reality is that the origin of the magnetic field of the active region is still unclear at present. One possibility is that the magnetic field of the flux rope or flux tube is forming while they are rising under the subsurface. Thus, a greater number of episodes ought to be considered in the numerical simulation. After all, we conclude that the simple active region does not need too much initial twist in the intrinsic magnetic field for its emergence.
\begin{acknowledgements}
      We appreciate the referee's careful reading of the manuscript and many constructive comments, which helped greatly in improving the paper.  SDO is a mission of NASA's Living With a Star Program. The authors are indebted to the SDO, NVST teams for providing the data. This work is supported by the National Key R\&D Program of China (2019YFA0405000), the National Science Foundation of China (NSFC) under grant numbers 12003064, 11873087, 11803085, Yunnan Science Foundation of China (2019FD085), the CAS ``Light of West China" Program under number Y9XB018001, the Open Research Program of the Key Laboratory of Solar Activity of Chinese Academy of Sciences (grant No.KLSA202014),  the Yunnan Talent Science Foundation of China (2018FA001), the Yunnan Science Foundation for Distinguished Young Scholars under No. 202001AV070004, and the Key Research and Development Project of Yunnan Province under number 202003AD150019..
\end{acknowledgements}


\begin{thebibliography}{}
\bibitem[Avallone \& Sun(2020)]{ava20} Avallone, E.~A. \& Sun, X.\ 2020, \apj, 893, 123
\bibitem[Babcock(1961)]{bab61} Babcock, H.~W.\ 1961, \apj, 133, 572
\bibitem[Berger, \& Field(1984)]{ber84} Berger, M.~A., \& Field, G.~B.\ 1984, Journal of Fluid Mechanics, 147, 133
\bibitem[Bernasconi et al.(2002)]{ber02} Bernasconi, P.~N., Rust, D.~M., Georgoulis, M.~K., et al.\ 2002, \solphys, 209, 119
\bibitem[Borrero et al.(2011)]{bor11} Borrero, J.~M., Tomczyk, S., Kubo, M., et al.\ 2011, \solphys, 273, 267
\bibitem[Cai et al.(2019)]{cai19} Cai, Q., Shen, C., Ni, L., et al.\ 2019, Journal of Geophysical Research (Space Physics), 124, 9824
\bibitem[Calabretta, \& Greisen(2002)]{cal02} Calabretta, M.~R., \& Greisen, E.~W.\ 2002, \aap, 395, 1077
\bibitem[Centeno(2012)]{cen12} Centeno, R.\ 2012, \apj, 759, 72
\bibitem[Centeno et al.(2014)]{cen14} Centeno, R., Schou, J., Hayashi, K., et al.\ 2014, \solphys, 289, 3531
\bibitem[Chen \& Shibata(2000)]{che00} Chen, P.~F., \& Shibata, K.\ 2000, \apj, 545, 524
\bibitem[Chen et al.(2018)]{che18} Chen, H., Yang, J., Yang, B., et al.\ 2018, \solphys, 293, 93
\bibitem[Cheung et al.(2010)]{che10} Cheung, M.~C.~M., Rempel, M., Title, A.~M., et al.\ 2010, \apj, 720, 233
\bibitem[Choudhuri \& Gilman(1987)]{cho87} Choudhuri, A.~R. \& Gilman, P.~A.\ 1987, \apj, 316, 788
\bibitem[Choudhuri et al.(1995)]{cho95} Choudhuri, A.~R., Schussler, M., \& Dikpati, M.\ 1995, \aap, 303, L29
\bibitem[Chintzoglou \& Zhang(2013)]{chi13} Chintzoglou, G. \& Zhang, J.\ 2013, \apjl, 764, L3
\bibitem[D'Silva \& Choudhuri(1993)]{dsi93} D'Silva, S. \& Choudhuri, A.~R.\ 1993, \aap, 272, 621
\bibitem[D{\'e}moulin, \& Berger(2003)]{dem03} D{\'e}moulin, P., \& Berger, M.~A.\ 2003, \solphys, 215, 203
\bibitem[D{\'e}moulin \& Pariat(2009)]{dem09} D{\'e}moulin, P. \& Pariat, E.\ 2009, Advances in Space Research, 43, 1013
\bibitem[Dikpati \& Gilman(2001)]{dik01} Dikpati, M. \& Gilman, P.~A.\ 2001, \apj, 559, 428
\bibitem[Fan(2001)]{fan01} Fan, Y.\ 2001, \apjl, 554, L111
\bibitem[Feynman \& Martin(1995)]{fey95} Feynman, J., \& Martin, S.~F.\ 1995, \jgr, 100, 3355
\bibitem[Fu \& Welsch(2016)]{fu16} Fu, Y. \& Welsch, B.~T.\ 2016, \solphys, 291, 383
\bibitem[Georgoulis et al.(2012)]{geo12} Georgoulis, M.~K., Titov, V.~S., \& Miki{\'c}, Z.\ 2012, \apj, 761, 61
\bibitem[Hagino, \& Sakurai(2004)]{hag04} Hagino, M., \& Sakurai, T.\ 2004, \pasj, 56, 831
\bibitem[Hale \& Nicholson(1925)]{hal25} Hale, G.~E. \& Nicholson, S.~B.\ 1925, \apj, 62, 270
\bibitem[Hoeksema et al.(2014)]{hoe14} Hoeksema, J.~T., Liu, Y., Hayashi, K., et al.\ 2014, \solphys, 289, 3483
\bibitem[Isobe et al.(2007)]{iso17} Isobe, H., Tripathi, D., \& Archontis, V.\ 2007, \apjl, 657, L53
\bibitem[Jiang et al.(2015)]{jia15} Jiang, J., Cameron, R.~H., \& Sch{\"u}ssler, M.\ 2015, \apjl, 808, L28
\bibitem[Knizhnik et al.(2018)]{kni18} Knizhnik, K.~J., Linton, M.~G., \& DeVore, C.~R.\ 2018, \apj, 864, 89
\bibitem[Knizhnik et al.(2021)]{kni20} Knizhnik, K.~J., Leake, J.~E., Linton, M.~G., et al.\ 2021, \apj, 907, 19
\bibitem[Krause \& Raedler(1980)]{kra80} Krause, F. \& Raedler, K.-H.\ 1980, Oxford, Pergamon Press, Ltd., 1980. 271 p.
\bibitem[Kosovichev \& Stenflo(2008)]{kos08} Kosovichev, A.~G. \& Stenflo, J.~O.\ 2008, \apjl, 688, L115
\bibitem[Kusano et al.(2002)]{kus02} Kusano, K., Maeshiro, T., Yokoyama, T., et al.\ 2002, \apj, 577, 501
\bibitem[Kutsenko et al.(2019)]{kut19} Kutsenko, A.~S., Abramenko, V.~I., \& Pevtsov, A.~A.\ 2019, \mnras, 484, 4393
\bibitem[Leka et al.(1996)]{lek96} Leka, K.~D., Canfield, R.~C., McClymont, A.~N., et al.\ 1996, \apj, 462, 547
\bibitem[Leka et al.(2009)]{lek09} Leka, K.~D., Barnes, G., Crouch, A.~D., et al.\ 2009, \solphys, 260, 83
\bibitem[Leighton(1964)]{lei64} Leighton, R.~B.\ 1964, \apj, 140, 1547
\bibitem[Li \& Ulrich(2012)]{li12} Li, J. \& Ulrich, R.~K.\ 2012, \apj, 758, 115
\bibitem[Li(2018)]{li18} Li, J.\ 2018, \apj, 867, 89
\bibitem[Luoni et al.(2011)]{luo11} Luoni, M.~L., D{\'e}moulin, P., Mandrini, C.~H., et al.\ 2011, \solphys, 270, 45
\bibitem[Liu \& Schuck(2012)]{liu12} Liu, Y., \& Schuck, P.~W.\ 2012, \apj, 761, 105
\bibitem[Liu et al.(2014)]{liu14} Liu, Y., Hoeksema, J.~T., Bobra, M., et al.\ 2014, \apj, 785, 13
\bibitem[Liu et al.(2017)]{liu17} Liu, Y., Sun, X., T{\"o}r{\"o}k, T., et al.\ 2017, \apjl, 846, L6
\bibitem[Liu \& Zhang(2006)]{liu06} Liu, J. \& Zhang, H.\ 2006, \solphys, 234, 21
\bibitem[Liu et al.(2014)]{liu14b} Liu, Z., Xu, J., Gu, B.-Z., et al.\ 2014, Research in Astronomy and Astrophysics, 14, 705-718
\bibitem[Magara(2001)]{mag01} Magara, T.\ 2001, \apj, 549, 608
\bibitem[Magara, \& Longcope(2003)]{mag03} Magara, T., \& Longcope, D.~W.\ 2003, \apj, 586, 630
\bibitem[Matsumoto et al.(1993)]{mat93} Matsumoto, R., Tajima, T., Shibata, K., et al.\ 1993, \apj, 414, 357
\bibitem[McClintock et al.(2014)]{mcc14} McClintock, B.~H., Norton, A.~A., \& Li, J.\ 2014, \apj, 797, 130
\bibitem[Melrose(1991)]{mel91} Melrose, D.~B.\ 1991, \apj, 381, 306
\bibitem[Metcalf(1994)]{met94} Metcalf, T.~R.\ 1994, \solphys, 155, 235
\bibitem[Murray et al.(2006)]{mur06} Murray, M.~J., Hood, A.~W., Moreno-Insertis, F., et al.\ 2006, \aap, 460, 909
\bibitem[Nandy(2006)]{nan06} Nandy, D.\ 2006, Journal of Geophysical Research (Space Physics), 111, A12S01
\bibitem[Okamoto et al.(2009)]{oka09} Okamoto, T.~J., Tsuneta, S., Lites, B.~W., et al.\ 2009, \apj, 697, 913
\bibitem[Parker(1955)]{par55} Parker, E.~N.\ 1955, \apj, 122, 293
\bibitem[Parker(1996)]{par96} Parker, E.~N.\ 1996, \apj, 471, 485
\bibitem[Pariat et al.(2005)]{par05} Pariat, E., D{\'e}moulin, P., \& Berger, M.~A.\ 2005, \aap, 439, 1191
\bibitem[Pesnell et al.(2012)]{pes12} Pesnell, W.~D., Thompson, B.~J., \& Chamberlin, P.~C.\ 2012, \solphys, 275, 3
\bibitem[Pevtsov et al.(1994)]{pev94} Pevtsov, A.~A., Canfield, R.~C., \& Metcalf, T.~R.\ 1994, \apjl, 425, L117
\bibitem[Pevtsov et al.(1995)]{pev95} Pevtsov, A.~A., Canfield, R.~C., \& Metcalf, T.~R.\ 1995, \apjl, 440, L109
\bibitem[Poisson et al.(2015)]{poi15} Poisson, M., Mandrini, C.~H., D{\'e}moulin, P., et al.\ 2015, \solphys, 290, 727
\bibitem[Rempel \& Cheung(2014)]{rem14} Rempel, M. \& Cheung, M.~C.~M.\ 2014, \apj, 785, 90
\bibitem[Schlichenmaier et al.(2010)]{sch10} Schlichenmaier, R., Bello Gonz{\'a}lez, N., Rezaei, R., et al.\ 2010, Astronomische Nachrichten, 331, 563
\bibitem[Schou et al.(2012)]{sch12} Schou, J., Scherrer, P.~H., Bush, R.~I., et al.\ 2012, \solphys, 275, 229
\bibitem[Schuck(2006)]{sch06} Schuck, P.~W.\ 2006, \apj, 646, 1358
\bibitem[Schuck(2008)]{sch08} Schuck, P.~W.\ 2008, \apj, 683, 1134
\bibitem[Seehafer(1990)]{see90} Seehafer, N.\ 1990, \solphys, 125, 219
\bibitem[Sokoloff et al.(2015)]{sok15} Sokoloff, D., Khlystova, A., \& Abramenko, V.\ 2015, \mnras, 451, 1522
\bibitem[Stein et al.(2011)]{ste11} Stein, R.~F., Lagerfj{\"a}rd, A., Nordlund, {\r{A}}., et al.\ 2011, \solphys, 268, 271
\bibitem[Stenflo \& Kosovichev(2012)]{ste12} Stenflo, J.~O. \& Kosovichev, A.~G.\ 2012, \apj, 745, 129
\bibitem[Strous et al.(1996)]{str96} Strous, L.~H., Scharmer, G., Tarbell, T.~D., et al.\ 1996, \aap, 306, 947
\bibitem[Thornton \& Parnell(2011)]{tho11} Thornton, L.~M., \& Parnell, C.~E.\ 2011, \solphys, 269, 13
\bibitem[Tian et al.(2018)]{tia18} Tian, H., Zhu, X., Peter, H., et al.\ 2018, \apj, 854, 174
\bibitem[Toriumi \& Yokoyama(2010)]{tor10} Toriumi, S. \& Yokoyama, T.\ 2010, \apj, 714, 505
\bibitem[Toriumi et al.(2011)]{tor11} Toriumi, S., Miyagoshi, T., Yokoyama, T., et al.\ 2011, \pasj, 63, 407
\bibitem[T{\"o}r{\"o}k et al.(2014)]{tor14} T{\"o}r{\"o}k, T., Leake, J.~E., Titov, V.~S., et al.\ 2014, \apjl, 782, L10
\bibitem[van Driel-Gesztelyi \& Green(2015)]{van15} van Driel-Gesztelyi, L., \& Green, L.~M.\ 2015, Living Reviews in Solar Physics, 12, 1
\bibitem[Vemareddy(2015)]{vem15} Vemareddy, P.\ 2015, \apj, 806, 245
\bibitem[Vemareddy \& D{\'e}moulin(2017)]{vem17} Vemareddy, P., \& D{\'e}moulin, P.\ 2017, \aap, 597, A104
\bibitem[Wang \& Abramenko(2000)]{wan00} Wang, T., \& Abramenko, V.~I.\ 2000, \aap, 357, 1056
\bibitem[Wang \& Sheeley(1989)]{wan89} Wang, Y.-M. \& Sheeley, N.~R.\ 1989, \solphys, 124, 81
\bibitem[Wang \& Sheeley(1991)]{wan91} Wang, Y.-M. \& Sheeley, N.~R.\ 1991, \apj, 375, 761
\bibitem[Wang et al.(2018)]{wan18} Wang, J., Yan, X., Qu, Z., et al.\ 2018, \apj, 863, 180
\bibitem[Wang et al.(2019)]{wan19} Wang, J., Yan, X., Guo, Q., et al.\ 2019, \mnras, 488, 3794
\bibitem[Wiegelmann et al.(2014)]{wie14} Wiegelmann, T., Thalmann, J.~K., \& Solanki, S.~K.\ 2014, \aapr, 22, 78
\bibitem[Xiang et al.(2016)]{xia16} Xiang, Y.-. yuan ., Liu, Z., \& Jin, Z.-. yu .\ 2016, \na, 49, 8
\bibitem[Yan et al.(2017)]{yan17} Yan, X.~L., Jiang, C.~W., Xue, Z.~K., et al.\ 2017, \apj, 845, 18
\bibitem[Yan et al.(2020)]{yan20} Yan, X., Liu, Z., Zhang, J., et al.\ 2020, Science in China E: Technological Sciences, 63, 1656
\bibitem[Yang et al.(2009)]{yan09} Yang, S., Zhang, H., \& B{\"u}chner, J.\ 2009, \aap, 502, 333
\bibitem[Young et al.(2018)]{you18} Young, P.~R., Tian, H., Peter, H., et al.\ 2018, \ssr, 214, 120
\bibitem[Zhukova et al.(2020)]{zhu20} Zhukova, A., Khlystova, A., Abramenko, V., et al.\ 2020, \solphys, 295, 165
\end{thebibliography}
\end{document}